\documentclass[final]{svjour2}
\usepackage{graphicx}
\usepackage{rotating}
\usepackage{amssymb}
\usepackage{mathptmx}
\usepackage[numbers,sort&compress]{natbib}
\usepackage{lipsum}
\usepackage{subfig}
\usepackage{sidecap}
\usepackage{wasysym}
\usepackage{bm}
\usepackage{tabularx}
\usepackage[hang,flushmargin,bottom]{footmisc} 

\makeatletter
\newcommand\subparagraph{%
  \@startsection{subparagraph}{5}
  {\parindent}
  {3.25ex \@plus 1ex \@minus .2ex}
  {-1em}
  {\normalfont\normalsize\bfseries}}
\makeatother
\makeatletter
\def\@maketitle{\newpage
\normalfont
\vbox to0pt{\if@twocolumn\vskip-39pt\else\vskip-49pt\fi
\nointerlineskip
\makeheadbox\vss}\nointerlineskip
\vbox to 0pt{\offinterlineskip\rubricwidth=\columnwidth
\vskip-12.5pt
\if@twocolumn\else % one column journal
   \divide\rubricwidth by144\multiply\rubricwidth by89 % perform golden section
   \vskip-\topskip
\fi
\hrule\@height0.35mm\noindent
\advance\fboxsep by.25mm
\global\advance\rubricwidth by0pt
\rubric
\vss}\vskip-42pt % skip between Journal header and list of authors
\if@twocolumn\else
 \gdef\footnoterule{%
  \kern-3\p@
  \hrule\@width\columnwidth  %rubricwidth
  \kern2.6\p@}
\fi
 \setbox\authrun=\vbox\bgroup
     \hrule\@height 9mm\@width0\p@
     \pretolerance=10000
     \rightskip=0pt plus 4cm
    \nothanksmarks
    {\authorfont
    \setbox0=\vbox{\setcounter{auth}{1}\def\and{\stepcounter{auth} }%
                   \hfuzz=2\textwidth\def\thanks##1{}\@author}%
    \setcounter{footnote}{0}%
    \global\value{inst}=\value{auth}%
    \setcounter{auth}{1}%
    \if@twocolumn
       \rightskip43mm plus 4cm minus 3mm
    \else % one column journal
    \fi
\def\and{\unskip\nobreak\enskip{\boldmath$\cdot$}\enskip\ignorespaces}%
    \noindent\ignorespaces\@author\vskip3.62pt} % space just after author block
    {\LARGE\bfseries
     \noindent\ignorespaces
     \@title \par}\vskip 1.62pt\relax % space just after title
    \if!\@subtitle!\else
      {\large\bfseries
      \pretolerance=10000
      \rightskip=0pt plus 3cm
      \vskip-5pt
      \noindent\ignorespaces\@subtitle \par}\vskip 0.24pt
    \fi
    \small
    \if!\@dedic!\else
       \par
       \normalsize\it
       \addvspace\baselineskip
       \noindent\@dedic
    \fi
 \egroup % end of header box
 \@tempdima=\headerboxheight
 \advance\@tempdima by-\ht\authrun
 \unvbox\authrun
 \ifdim\@tempdima>0pt
    \vrule width0pt height\@tempdima\par
 \fi
 \noindent{\small\@date\vskip -2.24pt} % skip after date
 \global\@minipagetrue
 \global\everypar{\global\@minipagefalse\global\everypar{}}%
}
\makeatother
\usepackage{titlesec}
\let\subparagraph\relax % You don't want to use \subparagraph

\usepackage{titlesec}
\makeatletter
\journalname{Journal of Low Temperature Physics}
\titlespacing\section{0pt}{12pt plus 4pt minus 10pt}{0pt plus 2pt minus 10pt}
\titlespacing\subsection{0pt}{12pt plus 4pt minus 10pt}{0pt plus 2pt minus 10pt}
\titlespacing\subsubsection{0pt}{12pt plus 4pt minus 10pt}{0pt plus 2pt minus 10pt}
\bibpunct{}{}{,}{s}{}{,}
\newcommand*{\doi}[1]{DOI:#1}
\newcommand{\s}[1]{\S\ref{#1}}

\newcommand{\EE}{\emph{E\kern0.75ptE} }
\newcommand{\TT}{\emph{T\kern0.75ptT} }
\newcommand{\BB}{\emph{B\kern0.75ptB} }

\begin{document}
\setcitestyle{numbers,square}

\newcommand{\hdblarrow}{H\makebox[0.9ex][l]{$\downdownarrows$}-}
\title{Advanced ACTPol Cryogenic Detector Arrays and Readout}

\author{
S.W.~Henderson\textsuperscript{1}\kern-1.5pt \and
R.~Allison\textsuperscript{2}\kern-1.5pt \and    
J.~Austermann\textsuperscript{3}\kern-1.5pt \and 
T.~Baildon\textsuperscript{4}\kern-1.5pt \and    
N.~Battaglia\textsuperscript{5}\kern-1.5pt \and  
J.A.~Beall\textsuperscript{3}\kern-1.5pt \and    
D.~Becker\textsuperscript{3}\kern-1.5pt \and     
F.~De Bernardis\textsuperscript{1}\kern-1.5pt \and   
J.R.~Bond\textsuperscript{6}\kern-1.5pt \and         
E.~Calabrese\textsuperscript{5}\kern-1.5pt \and      
S.K.~Choi\textsuperscript{7}\kern-1.5pt \and         
K.P.~Coughlin\textsuperscript{4}\kern-1.5pt \and     
K.T.~Crowley\textsuperscript{7}\kern-1.5pt \and      
R.~Datta\textsuperscript{4}\kern-1.5pt \and          
M.J.~Devlin\textsuperscript{8}\kern-1.5pt \and       
S.M.~Duff\textsuperscript{3}\kern-1.5pt \and         
R.~D\"unner\textsuperscript{9}\kern-1.5pt \and       
J.~Dunkley\textsuperscript{2}\kern-1.5pt \and        
A.~van~Engelen\textsuperscript{6}\kern-1.5pt \and    
P.A.~Gallardo\textsuperscript{1}\kern-1.5pt \and     
E.~Grace\textsuperscript{7}\kern-1.5pt \and          
M.~Hasselfield\textsuperscript{5}\kern-1.5pt \and    
F.~Hills\textsuperscript{4}\kern-1.5pt \and          
G.C.~Hilton\textsuperscript{3}\kern-1.5pt \and       
A.D.~Hincks\textsuperscript{10}\kern-1.5pt \and      
R.~Hlo\^{z}ek\textsuperscript{5}\kern-1.5pt \and    
S.P.~Ho\textsuperscript{7}\kern-1.5pt \and          
J.~Hubmayr\textsuperscript{3}\kern-1.5pt \and       
K.~Huffenberger\textsuperscript{11}\kern-1.5pt \and 
J.P.~Hughes\textsuperscript{12}\kern-1.5pt \and     
K.D.~Irwin\textsuperscript{13}\kern-1.5pt \and      
B.J.~Koopman\textsuperscript{1}\kern-1.5pt \and     
A.B.~Kosowsky\textsuperscript{14}\kern-1.5pt \and   
D.~Li\textsuperscript{3,15}\kern-1.5pt \and         
J.~McMahon\textsuperscript{4}\kern-1.5pt \and       
C.~Munson\textsuperscript{4}\kern-1.5pt \and        
F.~Nati\textsuperscript{8}\kern-1.5pt \and          
L.~Newburgh\textsuperscript{16}\kern-1.5pt \and     
M.D.~Niemack\textsuperscript{1}\kern-1.5pt \and     
P.~Niraula\textsuperscript{7}\kern-1.5pt \and       
L.A.~Page\textsuperscript{7}\kern-1.5pt \and        
C.G.~Pappas\textsuperscript{7}\kern-1.5pt \and      
M.~Salatino\textsuperscript{7}\kern-1.5pt \and      
A.~Schillaci\textsuperscript{7,17}\kern-1.5pt \and  
B.L.~Schmitt\textsuperscript{8}\kern-1.5pt \and     
N.~Sehgal\textsuperscript{18}\kern-1.5pt \and       
B.D.~Sherwin\textsuperscript{19}\kern-1.5pt \and    
J.L.~Sievers\textsuperscript{20}\kern-1.5pt \and    
S.M.~Simon\textsuperscript{7}\kern-1.5pt \and       
D.N.~Spergel\textsuperscript{5}\kern-1.5pt \and 
S.T.~Staggs\textsuperscript{7}\kern-1.5pt \and      
J.R.~Stevens\textsuperscript{1}\kern-1.5pt \and     
R.~Thornton\textsuperscript{21}\kern-1.5pt \and     
J.~Van Lanen\textsuperscript{3}\kern-1.5pt \and     
E.M.~Vavagiakis\textsuperscript{1}\kern-1.5pt \and  
J.T.~Ward\textsuperscript{8}\kern-1.5pt \and        
E.J.~Wollack\textsuperscript{22}\kern-1.5pt         
}
\institute{\footnotesize
  \noindent\textsuperscript{1}Department of Physics, Cornell University, Ithaca, NY, USA 14853\\
  \noindent\textsuperscript{2}Sub-Department of Astrophysics, University of Oxford, Keble Road, Oxford, UK OX1 3RH\\
  \noindent\textsuperscript{3}NIST Quantum Devices Group, 325 Broadway Mailcode 817.03, Boulder, CO, USA 80305\\
  \noindent\textsuperscript{4}Department of Physics, University of Michigan, Ann Arbor, USA 48103\\
  \noindent\textsuperscript{5}Department of Astrophysical Sciences, Peyton Hall, Princeton University, Princeton, NJ USA 08544\\
  \noindent\textsuperscript{6}Canadian Institute for Theoretical Astrophysics, University of Toronto, Toronto, ON, Canada M5S 3H8\\
  \noindent\textsuperscript{7}Joseph Henry Laboratories of Physics, Jadwin Hall, Princeton University, Princeton, NJ, USA 08544\\
\email{swh76@cornell.edu}}

\maketitle

\begin{abstract}

Advanced ACTPol is a polarization-sensitive upgrade for the $6$~m aperture Atacama Cosmology Telescope (ACT), adding new frequencies and increasing sensitivity over the previous ACTPol receiver. 
In 2016, Advanced ACTPol will begin to map approximately half the sky in five frequency bands ($28$\textendash$230$~GHz).
Its maps of primary and secondary cosmic microwave background (CMB) anisotropies -- imaged in intensity and polarization 
at few arcminute-scale resolution -- will enable precision cosmological constraints and also a wide array of cross-correlation 
science that probes the expansion history of the universe and the growth of structure via gravitational collapse. 
To accomplish these scientific goals, the Advanced ACTPol receiver will be a significant upgrade to the ACTPol receiver, including four new multichroic arrays of cryogenic, feedhorn-coupled AlMn transition edge sensor (TES) polarimeters (fabricated on 150 mm diameter wafers); a system of continuously rotating meta-material silicon half-wave plates; and a new multiplexing readout architecture which uses superconducting quantum interference devices (SQUIDs) and time division to achieve a 64-row multiplexing factor. 
Here we present the status and scientific goals of the Advanced ACTPol instrument, emphasizing the design and implementation of the Advanced ACTPol cryogenic detector arrays.


\end{abstract}

\pagebreak
%%% spill-over affiliations
\let\thefootnote\relax\footnotetext{\noindent\textsuperscript{8}Department of Physics and Astronomy, University of Pennsylvania, 209 South 33rd Street, Philadelphia, PA, USA 19104}
\let\thefootnote\relax\footnotetext{\noindent\textsuperscript{9}Departamento de Astronom\'{i}a y Astrof\'{i}sica, Pontic\'{i}a Universidad Cat\'{o}lica, Casilla 306, Santiago 22, Chile}
\let\thefootnote\relax\footnotetext{\noindent\textsuperscript{10}Department of Physics and Astronomy, University of British Columbia, Vancouver, BC, Canada V6T 1Z4}
\let\thefootnote\relax\footnotetext{\noindent\textsuperscript{11}Department of Physics, Florida State University, Tallahassee, FL, USA 32306}
\let\thefootnote\relax\footnotetext{\noindent\textsuperscript{12}Department of Physics and Astronomy, Rutgers, The State University of New Jersey, Piscataway, NJ USA 08854-8019}
\let\thefootnote\relax\footnotetext{\noindent\textsuperscript{13}Department of Physics, Stanford University, Stanford, CA, USA 94305-4085}
\let\thefootnote\relax\footnotetext{\noindent\textsuperscript{14}Department of Physics and Astronomy, University of Pittsburgh, Pittsburgh, PA, USA 15260}
\let\thefootnote\relax\footnotetext{\noindent\textsuperscript{15}SLAC National Accelerator Laboratory, 2575 Sandy Hill Road, Menlo Park, CA 94025}
\let\thefootnote\relax\footnotetext{\noindent\textsuperscript{16}Dunlap Institute, University of Toronto, 50 St. George St., Toronto, ON M5S 3H4}
\let\thefootnote\relax\footnotetext{\noindent\textsuperscript{17}Sociedad Radiosky Asesor\'{i}as de Ingenier\'{i}a Limitada Lincoy\'{a}n 54, Depto 805 Concepci\'{o}n, Chile}

\section{\label{sec:intro}Introduction}
\vspace{1em}
A worldwide experimental effort is underway to map the polarization of the CMB (see, for example~\cite{niemack2010,BICEP2a,Keck2012,Polarbear2010,SPTPol2012,EBEX2008,SPIDER2013,Planck2010,CLASS2014}).
These measurements will provide constraints on the parameters of the standard $\Lambda$CDM model that are independent of those
obtained from the CMB temperature alone.  Moreover, the parameters may be more tightly constrained by the former due to the higher 
contrast of the acoustic features in polarization compared to astrophysical foregrounds~\cite{galli2014,calabrese2014}. 
CMB polarization also probes multiple types of physics beyond the standard model,
including measuring the sum of the neutrino masses ($\sum m_{\nu}$) and constraining the tensor-to-scalar ratio of primordial fluctuations ($r$).
The polarization anisotropies in the CMB are more than an order of magnitude fainter than the temperature anisotropies, 
requiring large arrays of detectors integrating over long periods of time to detect them.  Optimized low-temperature detectors are photon-background
limited over the CMB frequency bands accessible from the ground, making them well-suited to these observations.
Substantial progress has been made in this field in the last two years, with ground-based experiments reporting 
the first high-significance detections of numerous new signals including so-called ``B-mode'' polarization at both small 
(sub-degree)~\cite{Hanson2013} and large (degree)~\cite{BICEP2014a} angular scales.  Current ground-based efforts include instruments such as the Atacama Cosmology Telescope 
Polarization-sensitive receiver (ACTPol)~\cite{niemack2010}, Polarbear~\cite{Polarbear2010}, SPTPol~\cite{SPTPol2012}, BICEP2~\cite{BICEP2a}, 
Keck-array~\cite{Keck2012}, and CLASS~\cite{CLASS2014}.  
This paper describes Advanced ACTPol (AdvACT), an upgrade to the ACTPol receiver.
ACTPol has demonstrated near-photon-background limited operation in its published
scientific results from its first season~\cite{naess2014,Madhavacheril2015,Allison2015,vanEngelen2015,grace2014}, 
and has presented preliminary results from its second and third seasons~\cite{hoLTD16,dattaLTD16}.  

In \s{sec:science} we present the scientific objectives of the AdvACT upgrade and in \s{sec:instrument} we
describe the instrumental upgrades designed to achieve these objectives, and we conclude in \s{sec:status} 
with a discussion of the status and future of the experiment.

\section{\label{sec:science}Scientific Objectives and Motivation}
\vspace{1em}
With AdvACT we plan to map approximately half of the microwave sky in five frequency bands spanning from 28 GHz to 230 GHz with four new arrays of detectors.
In addition to extended spectral coverage, AdvACT will have excellent 
angular resolution (expected: $1.4'$ at $150$~GHz, $7.1'$ at $28$~GHz), and increased polarization and temperature sensitivity 
as a result of roughly doubling the number of detectors per array in the mid- and high-frequency optical bands relative to ACTPol.
Polarization systematics and low-frequency receiver noise will be controlled
with the use of ambient temperature continously rotating metamaterial silicon half-wave plates (HWPs), which will modulate the incoming polarized
signals at ${\sim}8$~Hz, improving access to the large angular scale signals expected from inflation.
\begin{SCfigure}
\centering
\includegraphics[width=0.55\linewidth,keepaspectratio]{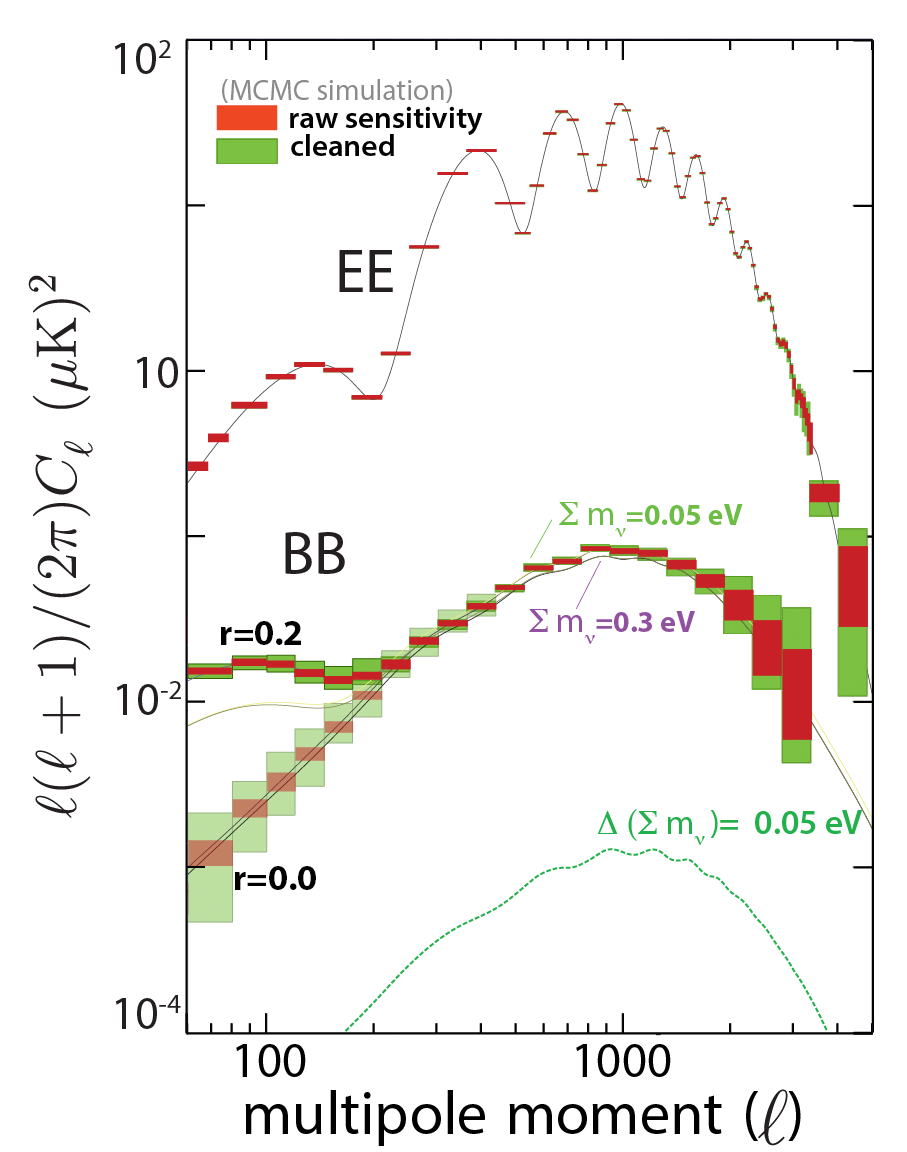} % 
\caption{Forecast AdvACT errors on \EE and \BB spectra, compared to $\Lambda$CDM predictions, for several values of $r$ and $\sum m_{\nu}$.  
The red error bars indicate the raw sensitivity.  The green error bars indicate an estimate for the degradation due to cleaning foregrounds 
that approximate Planck data (3x dustier than the nominal PSM), based on fitting a simple blackbody CMB plus power-law dust and synchrotron model
in each degree-scale patch.  This foreground projection is estimated to increase the map noise level to $\approx10-12$~$\mu$K-arcmin, although in practice the noise increase is expected to have scale dependence.  (Color figure online).}
\label{fig:forecast}
\end{SCfigure}

\let\thefootnote\relax\footnotetext{\noindent\textsuperscript{18}Physics and Astronomy Department, Stony Brook University, Stony Brook, NY USA 11794}
\let\thefootnote\relax\footnotetext{\noindent\textsuperscript{19}Berkeley Center for Cosmological Physics, University of California, Berkeley, CA, USA 94720}
\let\thefootnote\relax\footnotetext{\noindent\textsuperscript{20}Astrophysics and Cosmology Research Unit, School of Mathematics, Statistics and Computer Science, University of KwaZulu-Natal, Durban 4041, South Africa}
\let\thefootnote\relax\footnotetext{\noindent\textsuperscript{21}Department of Physics, West Chester University of Pennsylvania, West Chester, PA, USA 19383}
\let\thefootnote\relax\footnotetext{\noindent\textsuperscript{22}NASA Goddard Space Flight Center, Greenbelt, MD 20771 USA}

With these advances in sensitivity and systematics, we anticipate that measurements with AdvACT will substantially improve our understanding of cosmology,
nearly halving the uncertainty on many basic cosmological parameters including the baryon and matter densities, $\Omega_{b}h^{2}$ and $\Omega_{c}h^{2}$, as well as the spectral index $n_{s}$.  
The wide spectral coverage from the AdvACT arrays, combined with existing {\it Planck} $353$~GHz maps, has been chosen for efficiently detecting and removing synchrotron and dust 
foregrounds, targeting a cosmic variance-limited measurement of the primordial CMB polarization 
out to $\ell=2000$.
The large AdvACT survey region has been selected to increase the sensitivity for inflationary gravity wave searches
and permit measurements of the isotropy, frequency spectrum, and scale dependance
of any detected signal.  
Fig.~\ref{fig:forecast} shows the predicted errors on \emph{E\kern0.75ptE}- and \emph{B\kern0.75ptB}-mode power spectra expected for the complete AdvACT campaign, as compared with the spectra expected for different values of $r$ and $\sum m_{\nu}$.  
The red boxes show projected errors on the power spectra based on the predicted AdvACT raw instrument sensitivity, while the green boxes show projections after using the wide spectral coverage of AdvACT in conjunction with Planck measurements at $353$~GHz to remove dust foregrounds three times larger than in the Planck Sky Model (PSM)~\cite{Delabrouille2013}.

The measurements with AdvACT of secondary CMB anisotropies such as the thermal and
kinematic Sunyaev-Zel'dovich effects (tSZ and kSZ)~\cite{Carlstrom2002,Mueller2015} and gravitational lensing~\cite{smith2009}, will map the dark matter 
distribution, and potentially provide a high signal-to-noise measurement of the sum of the neutrino masses.  In addition, AdvACT precision \EE data will provide 
an independent measurement of the primordial \TT spectrum, allowing for a detection of the homogenous kSZ effect at ${>}10\sigma$ in AdvACT \TT data~\cite{calabrese2014}.

AdvACT benefits from extensive overlap with a large number of other existing and planned surveys, including the Large Synoptic Survey Telescope (LSST)~\cite{Ivezic2008LTD16}, as well as other 
optical surveys like HSC~\cite{HSC2006} and DES~\cite{DES2012}, X-ray surveys like eROSITA~\cite{eROSITA2012LTD16}, and spectroscopic surveys like SDSS-III~\cite{SDSIII2013} and DESI~\cite{DESI2013LTD16}.  This enables 
a wide array of cross-correlation science probing different epochs and is projected to result in the detection of ${>}10,000$ galaxy clusters at $99\%$ purity via their tSZ signatures and ${\sim}10,000$ high redshift, lensed, millimeter sources.  

\begin{table}[t]
\begin{center}
\label{table:instrument}
\begin{tabular}{|c|c|c|c|c|c|}
\hline
Detector & Center Freq. & Width  &  \# TES  & Projected Map Noise         & beam size \\ 
Array    & [GHz]        & [GHz]  &  --      & [$\mu$K-arcmin]             & [arcmin] \\ 
\hline
LF       &  28          &   6    &    88    & 80                          & 7.1  \\
LF       &  41          &  19    &    88    & 70                          & 4.8  \\
MF       &  90          &  39    &  1712    & 8                           & 2.2  \\
MF/HF    & 150          &  41    &  2718    & 7                           & 1.4  \\
HF       & 230          & 100    &  1006    & 25                          & 0.9  \\
\hline
\end{tabular}
\vspace{0.03in}
\caption{\label{tab:instdets}\small  Receiver parameters and projected sensitivity.
In total, AdvACT will deploy four new multichroic arrays; a high frequency (HF) array operating at $150$ and $230$~GHz, 
two medium frequency (MF) arrays operating at $90$ and $150$~GHz, and 
a low frequency (LF) array operating at $28$ and $41$~GHz.  
Map noise levels are based on three full years of observations, and are given in CMB temperature units (multiply by $\sqrt2$ for
polarization).
}
\end{center}
\vspace{-0.25in}
\end{table}%

\section{\label{sec:instrument}Instrument Overview}
\vspace{1em}

AdvACT will be an upgrade of ACTPol's three existing detector arrays and their optics, staged over three years.
The AdvACT receivers will be
deployed on the existing Atacama Cosmology Telescope~\cite{swetz2011}, located at 5190~m elevation in Parque Astron\'omico Atacama in northern Chile.  
AdvACT will reuse ACTPol's dilution refrigerator cryostat, which enables continuous ${\simeq}100$~mK observation.  The ACT telescope focuses sub-mm 
radiation onto each of three optics tubes in the cryostat, each of which contains a chain of filters and cold silicon reimaging optics which 
focus the radiation onto monolithic silicon feedhorn arrays directly coupled to arrays of transition-edge sensor (TES) bolometers.
AdvACT will deploy four new multichroic arrays, as summarized in Table~\ref{tab:instdets}.
As described in the following sections, the densely packed AdvACT multichroic
detector arrays require significant upgrades to the receiver optics, detectors, and readout to obtain
the required bandwidth and polarization sensitivity.

\subsection{{\it Optics}}
\vspace{1em}

The AdvACT receiver is located at the Gregorian focus of the ACT telescope. Each multichroic optics tube will have an ambient temperature metamaterial silicon 
half-wave plate (HWP) located in front of the optics tube window, optimized for its multichroic array bandpass. The HWPs are rotated at $\sim2$~Hz on custom 
air-bearing rotors, modulating the incoming polarization at $\sim8$~Hz.
Each HWP will be comprised of a stack of silicon wafers with machined grooves 
which together behave as an achromatic, birefringent material.  The Atacama B-Mode Search (ABS)
has demonstrated a reduction of low frequency atmospheric noise amplitude by a factor of greater than $500$ using a sapphire HWP, with less than $0.1$\% leakage
of intensity to polarization~\cite{Kusaka2014}.  
The metamaterial silicon HWPs for AdvACT will be twice as birefringent as sapphire, resulting in substantially less loss.  The broad band configuration will consist of a stack of three rotated 
HWPs~\cite{EBEXHWP} sandwiched between two additional silicon layers with grooved metamaterial AR coatings.
\begin{figure}%
    \centering
    \subfloat{  
      \scriptsize
      \raisebox{1.1\height}{
        \begin{tabular}{|c|c|c|c|}
          \hline
          TES                        & 150A/B           & 230A/B \\
          \hline
          AASP5 $R_{n}$ (m$\Omega$)   & $7.0{\pm}0.5$    & $7.4{\pm}0.2$ \\
          AASP6 $R_{n}$ (m$\Omega$)   & $7.6{\pm}0.1$    & $7.5{\pm}0.6$ \\
          \hline
          \textbf{Target} $\mathbf{R_{n}}$ \textbf{(m}$\boldsymbol{\Omega}$\textbf{)}  & \multicolumn{2}{c|}{$\mathbf{8}$} \\ %$8$     & $8$ \\
          \hline
          %% /home/swh76/repo/Shawn/analysis/plot_Tc_per_wafer.py
          AASP5 $T_{c}$ (mK)          & $178{\pm}1$     & $177{\pm}1$ \\
          AASP6 $T_{c}$ (mK)          & $182{\pm}2$     & $182{\pm}2$ \\
          \hline
          \textbf{Target $\mathbf{T_{c}}$ (mK)}          & \multicolumn{2}{c|}{$\mathbf{160}$} \\%$160$            & $160$ \\
          \hline
          AASP5 $P_{sat}$ (pW)        & $12.7{\pm}0.9$       & $24.3{\pm}0.6$ \\
          AASP6 $P_{sat}$ (pW)        & $14.1{\pm}0.8$       & $28{\pm}3$ \\
          \hline
          \textbf{Target $\mathbf{P_{sat}}$ (pW)}       & $\mathbf{12.5}$             & $\mathbf{25}$ \\
          \hline
          AASP5 $G$ (pW/K)           & $247{\pm}16$     & $557{\pm}11$ \\
          AASP6 $G$ (pW/K)           & $262{\pm}16$     & $580{\pm}35$ \\
          \hline
          \textbf{Target G (pW/K)}   & $\textbf{268}$   & $\textbf{536}$ \\
          \hline
          Leg Width ($\mu$m)         & $14.5$           & $28.5$ \\
          \hline
        \end{tabular}
      }
    }%
    %\qquad
    \subfloat{{\includegraphics[height=6.25cm] {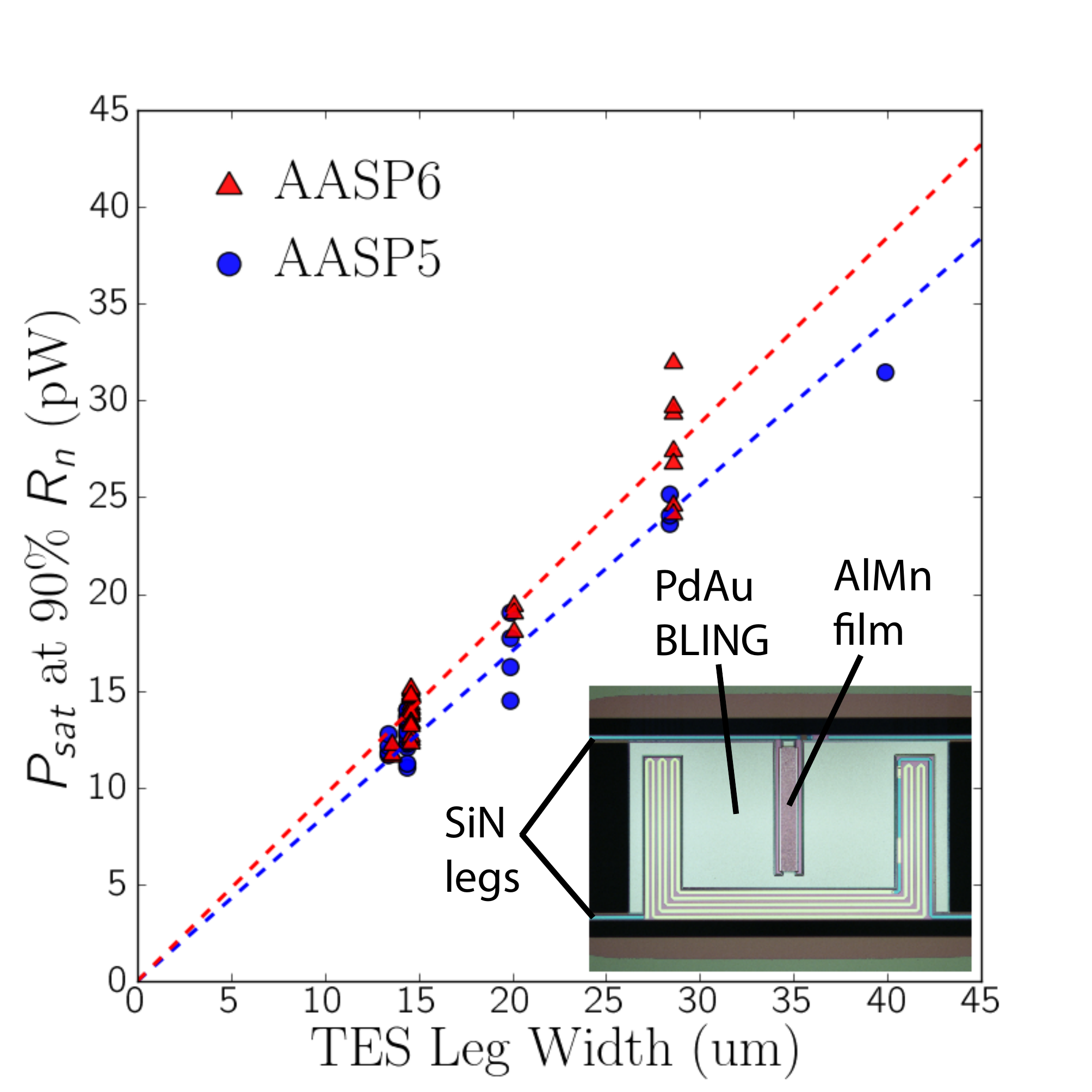} }} %
    \caption{{\it Left:} Measured average and RMS AdvACT detector parameters for many HF single pixels fabricated on two $75$~mm test wafers, AASP5 and AASP6, as compared with targeted values.  $R_{n}$ is the AlMn film normal resistance, $T_{c}$ the transition temperature, $P_{sat}$ the electrical bias power at $90$\% $R_{n}$, and $G$ the thermal conductance to the bath at $T_{c}$.  {\it Right:} $P_{sat}$ versus SiN leg width for typical HF $230$/$150$ and MF $90$~GHz devices.  {\it Right inset:} A microscope image of an AdvACT AlMn TES~\cite{liLTD16}.  (Color figure online.)}%
    \label{fig:hftes}%
\end{figure}

\begin{figure}%
    \centering
    \subfloat{{\includegraphics[height=3.65cm]{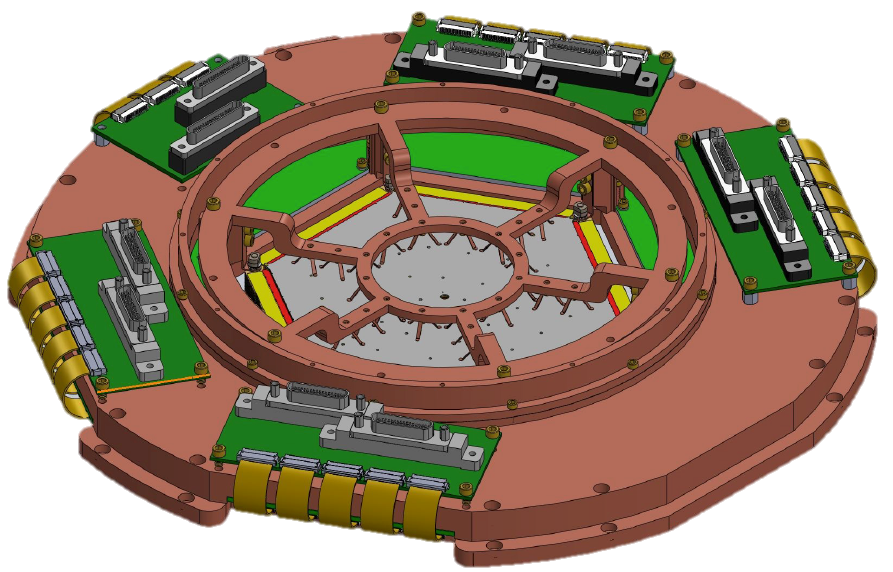} }}%
    %\qquad
    \subfloat{{\includegraphics[height=3.65cm]{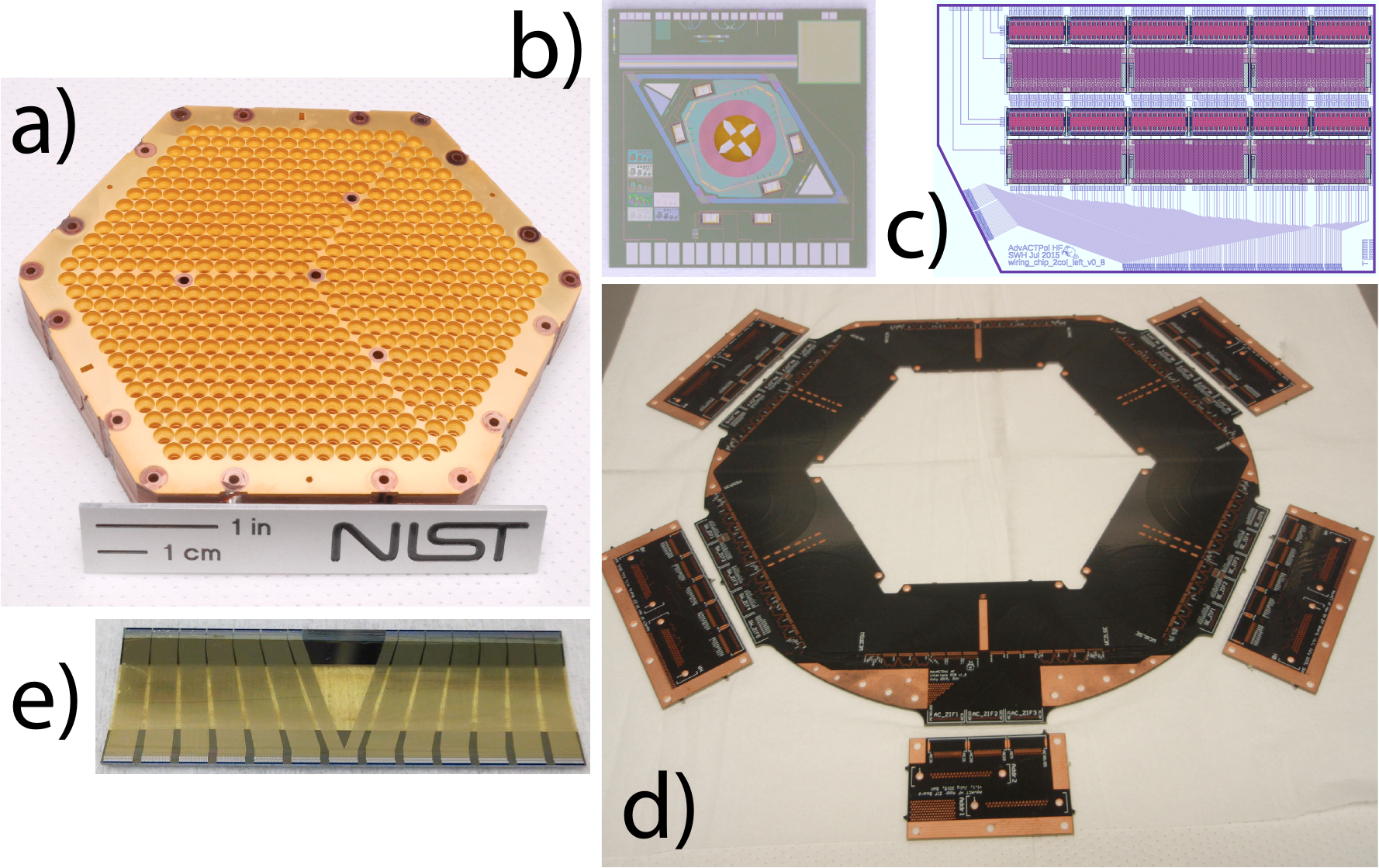} }}%
    \caption{{\it Left:} 3D model of the AdvACT HF array package.  The full array assembly is enclosed within a $100$~mK OHFC copper shield.  
      {\it Right a)} HF silicon micro-machined feedhorn array. 
      {\it b)} AdvACT HF test pixel.
      {\it c)} Layout for one of eight wiring chips that distribute TES signals from the array to the SQUID MUX, with MUX and detector bias chips overlaid.
      {\it d)} 6-layer PCBs that route signals from MCE to wiring chips. 
      {\it e)} Superconducting flex interface between HF array and main ring-shaped PCB.
      (Color figure online.)}%
    \label{fig:arraypkgandfeeds}%
\end{figure}

After the HWPs, light is focused by cryogenic silicon lenses onto a gold-plated silicon micro-machined platelet array of feedhorns, also optimized 
for the bandpass of its multichroic array.
The silicon lenses for each array have frequency matched, broadband AR metamaterial coatings consisting of three layers of grooves, proven
to have negligible in-band dielectric loss and reflection and excellent polarization symmetry~\cite{datta2013}.  The feedhorns are a new broad-band, spline-profiled smooth-wall design, optimized to produce
symmetric beams with high coupling-efficiency and low cross-polarization over the large bandwidths required by the AdvACT 
multichroic arrays.  Fig.~\ref{fig:arraypkgandfeeds} shows the first completed AdvACT feedhorn array.  

\subsection{{\it Detectors}}
\vspace{1em}
Each feed horn focuses light onto a single pixel, where it is coupled via planar superconducting microwave on-chip circuitry to four TES bolometers: 
two for each linear polarization state and two for each frequency band~\cite{datta2014}.
ACTPol has 
recently successfully deployed the first multichroic array for a CMB experiment, operating at $90$ and $150$~GHz~\cite{hoLTD16,dattaLTD16}.
Instead of tiling $75$~mm wafers to form each full array as in ACTPol, each AdvACT array will be fabricated on a single, 
$150$~mm diameter, $500$ $\mu$m thick silicon wafer, allowing for more efficient packing of the ACT focal plane and higher pixel densities~\cite{duffLTD16}.

The AdvACT TESes are fabricated from single-layer AlMn films instead of bilayer MoCu films as in ACTPol~\cite{deiker2004}.
Several rounds of single-pixel MF and HF prototypes have been successfully fabricated that meet the target specifications for 
AdvACT~\cite{liLTD16,austermannLTD16}.  
The results of extensive electrical and thermal characterization of HF devices 
from single pixels fabricated on two $75$~mm test wafers are compared to targeted parameters in Fig.~\ref{fig:hftes} 
and its table.

\subsection{{\it Readout}}
\vspace{1em}

Each TES in the AdvACT arrays will be DC voltage biased, and each array will be read out via a time division multiplexing (TDM)
scheme as implemented in the warm Multi-Channel Electronics (MCE)~\cite{Battistelli2008}.  The high TES density in the MF and HF arrays will 
require a multiplexing (MUX) factor, or number of detectors per readout channel, of 64:1, substantially larger than the
highest MCE MUX factor yet of 41:1 implemented on SCUBA2~\cite{holland2013}.  We have already demonstrated all the required 
electronics capabilites for a 64:1 MUX factor while reading out an 11 column by 44-row array of SQUIDs in the laboratory.

The arrays will be multiplexed through the MCE using a new TDM MUX architecture developed at NIST/Boulder with two SQUID stages 
and flux activated SQUID switches~\cite{beyer2008,dorieseLTD16}.  The current in each TES in the array is inductively coupled to its own 
first-stage SQUID (SQ1) and operated in a flux-locked loop that acts on the output of a second stage SQUID series array.
New cryogenic interfaces have been developed to connect each TES in the AdvACT arrays at $100$~mK into this cold multiplexing 
circuit.

At $100$~mK, signal pairs from each TES are carried to its own SQ1 via superconducting Nb traces on the wafer Al bonded to flexible 
ribbon cable containing superconducting Al traces. The voltage biasing is routed such that TESes observing at different optical frequencies 
are wired to different voltage bias lines and TESes from polarization pairs at one optical frequency for a given pixel are read out through 
the same $4$~K SQUID series array and $300$~K warm amplifier.
The SQ1s and associated wiring and addressing chips are located on a large printed circuit board (PCB) which surrounds the array. It contains routing and bondpads 
to the appropriate silicon chips for all necessary control lines: TES biases, SQUID biases and feedbacks, and switching currents.
In total, the AdvACT MF 
and HF arrays will each require in excess of $20,000$ aluminum wirebonds, and numerous new tools and procedures have been developed 
to ensure a robust and high yield integration~\cite{pappasLTD16}.  The entire AdvACT HF 
array mechanical and readout $100$~mK assembly is shown in Fig.~\ref{fig:arraypkgandfeeds}.

\section{\label{sec:status}Status}
\vspace{1em}
Observations with the first AdvACT array will begin in 2016, replacing one of the ACTPol arrays with the $150/230$~GHz AdvACT HF array.  
The staged deployment of the additional AdvACT arrays is planned for 2017 and 2018.

\vspace{-1em}
\begin{acknowledgements}
This work was supported by the U.S. National Science Foundation through awards
1312380 and 1440226.  The NIST authors would like to acknowledge the support
of the NIST Quantum Initiative. 
The development of multichroic detectors and lenses was supported by NASA
grants NNX13AE56G and NNX14AB58G.
The work of KPC, KTC, EG, BJK, CM, BLS, JTW, and SMS was supported by 
NASA Space Technology Research Fellowship awards. 
\end{acknowledgements}

\bibliography{apj-jour,advact}
\bibliographystyle{ltd16}

\end{document}